\documentclass[%
 reprint,
 amsmath,amssymb,
 aps,
prb,
floatfix,
]{revtex4-1}

\usepackage{graphicx}
\usepackage{dcolumn}
\usepackage{bm}

\begin{document}

\preprint{APS/123-QED}

\title{Non-Fermi Liquid Behaviour in the Three Dimensional Hubbard Model}

\author{Samuel Kellar}
 \affiliation{Department of Physics and Astronomy, Louisiana State University, Baton Rouge, Louisiana 70803, USA}
\author{Ka-Ming Tam}%
 \affiliation{Department of Physics and Astronomy, Louisiana State University, Baton Rouge, Louisiana 70803, USA}%

\date{\today}

\begin{abstract}
We present a numerical study on non-Fermi liquid behaviour of a three dimensional system. The Hubbard model in a cubic lattice is simulated by the dynamical cluster approximation, in particular the quasi-particle weight is calculated at finite dopings for a range of temperatures. Near the putative quantum critical point, we find evidence of a separatrix at a finite doping which separates the Fermi liquid from non-Fermi liquid as the doping increases. Our results suggest that a marginal Fermi liquid and possibly a quantum critical point should exist in the three dimensions interacting Fermi system. 
\end{abstract}

\maketitle


\section{\label{sec:level1}Introduction}

The theory of the Fermi liquid is an important milestone of condensed matter physics. \cite{Landau1956,Landau1957,Landau1959} It encapsulates almost all metallic interacting fermionic systems. The fundamental assumption is that the interacting system can be obtained by adiabatically turning on the interaction. All the quantum numbers of the non-interacting system remain intact, specifically the momentum remains a good quantum number for characterizing excitations. 

There are notable exceptions to the Fermi liquid which have been discovered over time. The most prominent is the one dimensional system, in which different quantum numbers emerge due to the spin-charge separation. \cite{Voit_1995} In general, it is rather difficult to violate the assumption of the Fermi liquid as dictated by the phase space restriction in the particle-hole diagrams. \cite{Shankar1994,abrikosov1963methods} A general idea is either the density of state or the interaction becomes singular resulting in a strong correction from these effects. 

Along this line of thought, a possible cause of non-Fermi liquid behavior could be the proximity of a quantum critical point. A quantum critical point leads to long wavelength fluctuations. \cite{Hertz_1976,Moriya_Kawatabe_1973a,Millis_1993,Moriya_Kawatabe_1973b} The effective coupling of the electrons due to such fluctuations could become singular. We recognize it is usually difficult to obtain clean experimental evidence for a quantum critical point. In particular, the metallic critical point studied in this paper is difficult to observe as the critical point may be preempted by other orderings such as superconducting pairing. \cite{Lederer_etal_2017} Remarkable progress has been made recently on Kondo lattice materials. \cite{Steppke933,paschen2004hall} 

A prominent example of non-Fermi liquid behaviour is the high temperature superconducting cuprate. \cite{hightc} It is a non-Fermi liquid immediately above the superconducting dome. This has been extensively studied over the past three decades, as it is widely believed that a key to understanding the high temperature superconductivity is understanding the mechanism of the non-Fermi liquid, often denoted as strange metal. \cite{Schlesinger_1990,Varma_1999,MFL_1989} A theory which captures a lot of the strange metal behaviors is the theory of the marginal Fermi liquid. \cite{MFL_1989} The key idea of the marginal Fermi liquid is that the quasi-particle damping is much reduced to the point that quasi-particle excitations can still be defined. 

For the Fermi liquid, the quasi-particle damping reflects the Lorentzian shape of the excitation spectral weight. That is the imaginary part of the self energy scales as the square of the energy, $\Im[\Sigma(\omega)] \sim \omega^2$. Strictly speaking, a logarithmic correction is acquired at two dimensions. \cite{Varma_etal_2002} For the three or higher dimensional case, it can be shown that that the Fermi liquid is stable against particle-hole excitations either from perturbation theory or modern functional normalization group. \cite{abrikosov1963methods,Shankar1994} For the marginal Fermi liquid, the self energy scales linearly with respect to the energy, $\Im[\Sigma(\omega)] \sim \omega$. This is the borderline case for which the coherent excitations can be defined. This theory naturally explains one of the most interesting characteristics of the strange metal--linear resistivity. 

Numerically studying the non-Fermi liquid is challenging as it involves interacting fermions at finite doping. The dynamical cluster approximation (DCA) has been used to demonstrate the existence of a quantum critical point and marginal Fermi liquid in the two dimensional Hubbard model. \cite{2dqcp} This work extends the study to the three dimensional system. This paper is organized as follows. In the section II, we describe the model and review the previous studies of non-Fermi liquid and quantum criticality of the Hubbard model at two dimensions by the DCA. We also describe the method in the section. The results of the quasi-particle weight and the estimate of the crossover temperature between Fermi liquid and marginal Fermi liquid are described in the section III. We then conclude and discuss possible future work for additional evidence corroborating the existence of the quantum critical point in the three dimensional Hubbard model.

\section{\label{sec:Theory}Model and Method}

\subsection{Model}
Our starting point is the Hubbard model
\begin{equation}
    H = -t \sum_{<{i,j}>,\sigma} (c^{\dagger}_{i\sigma} c_{j\sigma} + H.c.)+ U \sum_{i} n_{i\uparrow} n_{i\downarrow} - \mu \sum_{i\sigma} n_{i\sigma},
\end{equation}
where $c^{\dagger}_{i\sigma}$ and $c_{i\sigma}$ are the creation and annihilation operators for electrons at site $i$ with spin $\sigma$. $n_{i,\sigma}$ is the number operator for site $i$ of spin $\sigma$. $t$ is the hopping energy between nearest neighbors of a simple cubic lattice. $U$ is the on-site repulsive coupling. The chemical potential, $\mu$, sets the filling of the system.

Built upon the two dimensional Hubbard model theory the three dimensional model acts as a bridge between the dynamical mean field approximation and the cuprate superconductors modeled so well by the two dimensional Hubbard model \cite{Zhang_Rice_1988}. The three dimensional Hubbard model has primarily been studied in terms of the antiferromagnetic properties at half-filling \cite{Fuchs_etal_2011,Kent_etal_2005,Karchev_2013,Staudt_2000}. Recently, the doped antiferromagnetic critical point has also been studied by the dynamical vertex approximation \cite{Schafer_etal_2015,Schafer_etal_2017}.
This research intends to further explore the features of the metallic phase of a doped three dimensional Hubbard model, in particular for searching the putative marginal fermi liquid. The hopping, $t$, is set to $0.25$ and it is used to set the energy scale. The bare bandwidth is $W=12t=3$. The on-site interaction is set to $U=0.75W$.

We use the DCA to solve the Hubbard model. The DCA employs clusters on a periodic lattice embedded in a dynamical mean field \cite{DCA_2000,DCA_RMP}. The lattice may be tiled exactly for a cubic lattice when the number of cluster points is a perfect cube. That limits the choices available. In order to find an intermediate size a Betts lattice is employed \cite{Betts_1998}. We choose a sixteen site cluster to keep the computational time accessible while still including a reasonable number of points in the first Brillouin zone.

\subsection{DCA Cluster}
\begin{figure}
    \centering
    \includegraphics[width=0.4\textwidth]{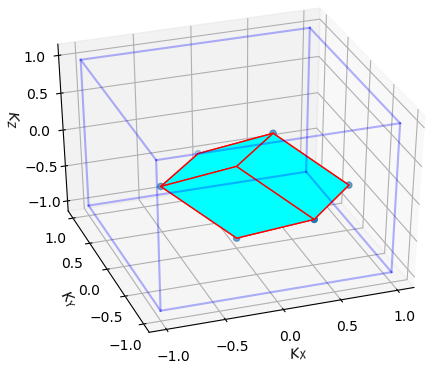}
    \caption{An example of the tiling used for the sixteen site cluster chosen for our simulation of the dynamical cluster approximation. The parallelepiped will be repeated to tile the entire Brillouin zone. The outline of the cube delineates the boundary of the first Brillouin zone. $K_x$, $K_y$ and $K_z$ are in the unit of $\pi$.}
    \label{fig:3dcluster}
\end{figure}

The sixteen site cluster has three vectors in momentum space defining the parallelepiped. They are
\begin{equation}
    a_1=(\frac{\pi}{2},\frac{\pi}{2},0)\quad a_2=(\frac{\pi}{2},-\frac{\pi}{2},-\frac{\pi}{2})\quad a_3=(-\frac{\pi}{2},\frac{\pi}{2},-\frac{\pi}{2}).\notag
\end{equation}
The first Brillouin zone is tiled by 16 of these parallelepipeds. The resulting 16 momentum points are not all unique. The location of the 8 unique points, in the first Brillouin zone, are
\begin{eqnarray}
    (-\frac{\pi}{2},-\frac{\pi}{2},\pi)\quad (-\frac{\pi}{2},-\frac{\pi}{2},0)\quad (0,0,\pi)\quad (-\frac{\pi}{2}, \frac{\pi}{2},\frac{\pi}{2})
     \notag \\ \quad (0,0,0) \quad (0,\pi,\frac{\pi}{2}) \quad (\pi,\pi,\pi) \quad (\pi,\pi,0). \notag
\end{eqnarray}
The cluster can be represented visually as shown in Fig. \ref{fig:3dcluster}. The periodic boundary conditions will be respected by the tiling as any point outside the zone will be mapped back inside at a symmetric point.

\subsection{CTQMC and Simulation Parameters}
The continuous time quantum Monte Carlo (CTQMC) was employed in order to solve the cluster impurity problem \cite{ctqmc_rmp,ctqmc}. In order to increase the precision of the Monte Carlo statistics the measurements should be uncorrelated. This depends upon sufficiently changing the sample being measured from its previous state. 
In order to  minimize the correlations between measurements after the system was warmed up the majority of proposed changes were flips of the vertices spins \cite{Mikelsons_Macridin_Jarrell_2009}. This occurred with a probability of about $70\%$ while additions and removals of vertices both occurred at a rate of about $15\%$. The DCA is employed on a high performance super computer. The simulation was run on 4200 CPUs in parallel. Each processor performed 250 measurements resulting in approximately $10^6$ measurements per iteration.  There were 100 proposed Monte Carlo steps between each measurement for each iteration of the DCA self-consistent cycle. Each CPU was warmed up independently with 10000 Monte Carlo steps.


\subsection{Locating the Fermi Surface}
The DCA gives the self-consistent values for the self energy at the sixteen points in the first Brillouin zone. The self energy is calculated for four hundred frequency points. In order to find the quasi-particle weight the Fermi surface is found along the $<1,1,1>$ direction. 
We identify the Fermi surface as the $\max(| \nabla n(\mathbf{k}) |)$, where $n(\mathbf{k})$ is the occupation number. Once the momentum of the Fermi surface in the chosen direction is identified that momentum point will be used to find the quasi-particle weight. 

In order to identify the Fermi surface a higher resolution of momentum space is required than is given by the allotted points. In order to achieve a sufficient resolution an interpolation through the points available is utilized after the self consistent DCA result is found. It is important to include information from all points due to the limited resolution in three dimensional space. Thus an inverse distance weighting \cite{inverseinterp} scheme was selected. 

The interpolation sums all available data in order to contribute to the interpolation of the new value. The sum is weighted such that the closest known data points have a much stronger weighting than the further points. The interpolation follows the formula
\begin{equation}
f(x)=
\begin{cases}
    \frac{\sum_{i} w_i(x) u_i}{\sum_i w_i(x)}, &d(x,x_i) \neq 0 \text{  for all } i \\ u_i, &d(x_i,x)=0 \text{ for some } i.
\end{cases}
\end{equation}
Where $i$ sums over all known points, $u_i$ are the known values. The location being interpolated is labeled $x$ and the location of the known values are $x_i$. $d(x,x_i)$ is the distance of the interpolated point to the known point, and $w_i$ is the weighting function given to each point. The weighting function is defined by
\begin{equation}
    w_i(x)=\frac{1}{d(x,x_i)^p}
\end{equation}
where $p$ is a parameter chosen to control the rate with which the weight drops off over distance. In this research a value of $6$ was used for $p$ though the interpolation results were found to be robust for a variety of $p$ values.

\subsection{Quasi-Particle Weight}
The lowest Matsubara frequency point of the self energy at the Fermi surface is then used to calculate the quasi-particle weight. Quasi-particle weight is defined in terms of the real frequency self energy. In order to relate the quasi-particle weight to the Matsubara self energy the following process is followed. The quasi-particle weight is related to the retarded self energy by,
\begin{equation}
    Z_{\mathbf{k}} = \frac{1}{1-\partial_\omega \left. \Re[\Sigma(\mathbf{k},\omega)]\right|_{\omega=0}}.
\end{equation}
Analytic continuation on the numerical data can be bypassed by taking the derivative of the Kramers-Kronig relation and then use the analytic continuation of the self energy $ 
    \Sigma(i \omega_n) = - \int \frac{d \omega}{\pi} \frac{\Im[\Sigma(\omega)]}{i \omega_n - \omega}$. \cite{Arsenault_etal_2012} 
    

The quasi-particle weight can then be approximated directly from the imaginary part of the self energy in Matsubara frequency, 
\begin{equation}
    Z_{\mathbf{k}} \approx \frac{1}{1-\Im[\Sigma(\mathbf{k},i\omega_0)]/ \omega_0}.
\end{equation}

\subsection{Fitting of the Quasi-Particle Weight}

We first consider the imaginary part of the real frequency self energy of both the Fermi liquid and the marginal Fermi liquid. \cite{2dqcp,abrikosov1963methods,MFL_1989} For the Fermi liquid, the imaginary part of the self energy has the form
\begin{equation}
\Im[\Sigma_{FL}(\omega)] = - \alpha \max\left(\omega^2,T^2\right),
\label{eq:SigmappFL}
\end{equation}
where $\alpha$ is a positive constant.

On the other hand the imaginary part of the marginal Fermi liquid self energy has the form
\begin{equation}
\Im[\Sigma_{MFL}(\omega)] = - \alpha \max\left( \left|\omega\right|,T\right)\,.
\label{eq:SigmappMFL}
\end{equation}

Near the putative quantum critical point, the single particle properties
of the model are observed to cross over from Fermi liquid to marginal Fermi liquid as the temperature
crosses $T_X$ and the frequency $\omega_X$. It was proposed that the self energy can be written in term of the so-called crossover form \cite{2dqcp}
\begin{equation}
\Im[\Sigma_{X}(\omega)] = \left\{ \begin{array}{c}
- \alpha \omega_X \max\left( \left|\omega\right|,T\right) \mbox{ for } |\omega|>\omega_X \mbox{ or } T>T_X \\
- \alpha \max\left(\omega^2,T^2\right) \mbox{ for } |\omega|<\omega_X \mbox{ and } T<T_X.
\end{array} \right.
\label{eq:SigmappX}
\end{equation}

From the analytic continuation, 
\begin{equation}
    \Sigma(\mathbf{k},i\omega_{n}) = - \int^{\omega_{c}}_{-\omega_{c}} \frac{d\omega \Im[\Sigma(\mathbf{k},\omega)]}{\pi(i\omega_n -\omega)},
\end{equation}
where $\omega_c$ is the cutoff at the order of the bandwidth. A crossover form for the self energy between the marginal Fermi liquid and the Fermi liquid states is found, \cite{2dqcp,functionalforms}
\begin{multline}
            \frac{\Im [\Sigma(i \omega_{0})]}{\omega_{0}}=\frac{-2 \alpha T}{\pi} \Theta(T_X-T) \left[\frac{\omega_X}{T}+0.066235 \right. \\ \left. {}-(0.308 \frac{\omega_X}{\pi T}  +\pi \tan^{-1} \frac{\omega_X}{\pi T})-\frac{\omega_X}{T}\ln (\frac{\omega_{X}^{2}+\pi^{2}T^{2}}{(1+\pi^2)T^2}) \right] \\
            +\omega_X\left[0.0981+\frac{1}{2}ln(\frac{\omega_{c}^{2}+\pi^{2}T^{2}}{(1+\pi^{2})T^{2}})\right].
            \label{eq:tx}
\end{multline}
The fitting parameters for this form are $\alpha$ a scaling parameter, $T_X$ the crossover temperature, $\omega_X$ the crossover frequency , and the cutoff $\omega_c$. With these parameters we can find a temperature ($T_X$) for which the system crosses from a marginal Fermi liquid to a Fermi liquid.

\section{Results}
The quasi-particle weight can be used to indicate whether or not a system is in a Fermi liquid state. In a Fermi liquid state the quasi-particle weight would be expected to have a finite value at zero temperature. While the true $T=0$ case cannot be simulated by quantum Monte Carlo solver for the DCA, the saturation of the quasi-particle weight at low temperatures allows for extrapolation to a finite value at $T=0$. In this way we determine a system to be in a Fermi liquid state. 
\begin{figure}
    \centering
    \includegraphics[width=.5\textwidth]{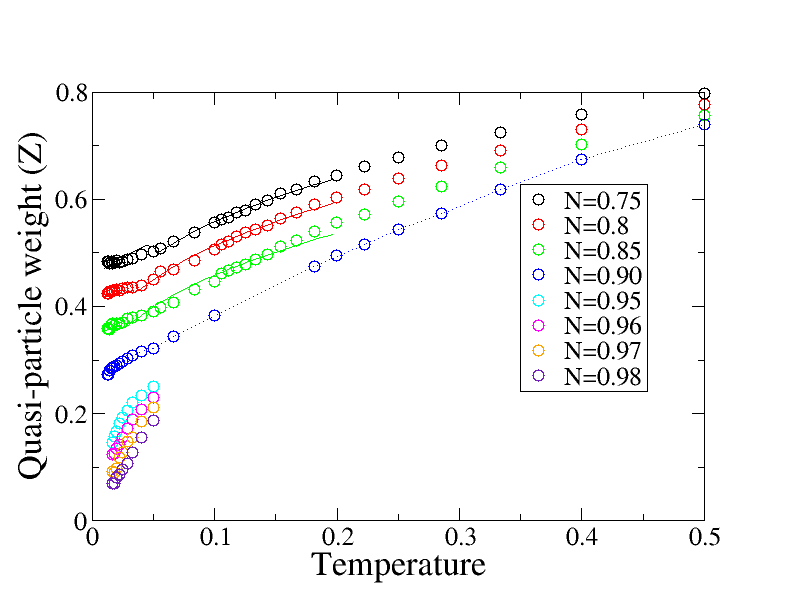}
    \caption{The quasi-particle weight as a function of filling, $N$. There is a clear separatrix between the behavior in highly doped systems and those near half-filling. The division occurs at about $N=0.95$. The dotted lines show the crossover form fit (see eq. \ref{eq:tx})  to the values of quasi-particle weight at the Fermi surface along the $<111>$ direction.}
    \label{fig:quasi}
\end{figure}
The quasi-particle weight of a non-Fermi liquid state has no residual quasi-particle weight at zero temperature. Again such a result must be extrapolated. The low temperature simulation will show a stark contrast to that of a Fermi liquid and will be readily distinguishable due to the rapid decrease in quasi-particle weight as temperature decreases.

The results included in Fig. \ref{fig:quasi} show a clear separation at a finite filling between Fermi liquid and non-Fermi liquid behavior. The simulation is run to the temperature of $T=0.0125$ below which the computational cost becomes prohibitively expensive. Above filling $N=0.95$ the quasi-particle weight tends towards 0 as the temperature decreases. The closer to half-filling the faster the quasi-particle weight asymptotically approaches 0. The cases where the system is in the Fermi liquid state show an asymptotic behavior towards a finite value for the quasi-particle weight.
 
At high temperature the system shows some evidence of being in a marginal Fermi liquid state. As such a fit of the quasi-particle weight for a crossover from the marginal Fermi liquid to Fermi liquid state was performed. The fit is shown as the lines for the various dopings above $N=0.95$ in Fig. \ref{fig:quasi}. 

\begin{figure}
    \centering
    \includegraphics[width=.49\textwidth]{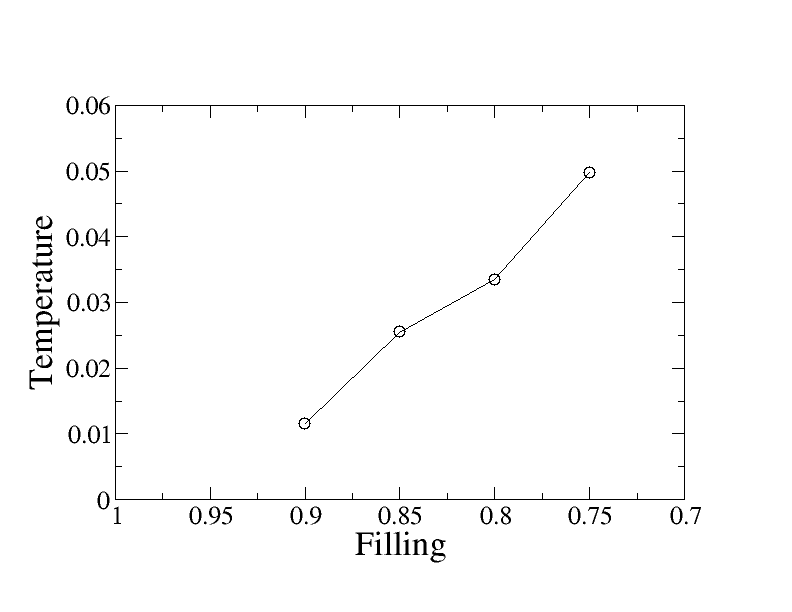}
    \caption{The crossover temperature, $T_X$, as a function of filling, $N$. The crossover temperature decreases as the filling increases. The decrease insinuates a critical filling at a finite doping before half-filling. This serves to highlight the possibility of a quantum critical point at a finite doping.}
    \label{fig:crossover}
\end{figure}

The fitted crossover form extracts the crossover temperature where the system is expected to begin behaving as a Fermi liquid. The results show that the crossover temperature decreases as the doping decreases. This is indicated in Fig. \ref{fig:crossover}. The crossover temperatures monotonically decrease and indicate a possible zero temperature crossover at a finite doping. This coincides well with the finding that the non-Fermi liquid state begins at $N=0.95$. This appears to be consistent with where the extrapolated filling would be at the zero temperature as shown in the crossover temperature graph in the fig. \ref{fig:crossover}. 


\section{Conclusion}
We study the quasi-particle weight of the doped three dimensional Hubbard model by the dynamical cluster approximation via the continuous time quantum Monte Carlo solver. We find that the imaginary part of the self energy fit into the crossover form, from the Fermi liquid to the marginal Fermi liquid, shows a monotonic decrease of crossover temperature as the filling increases. The putative critical filling nearly coincides with the filling in which quasi-particle weight decreases sharply as the temperature goes towards zero. It is tempting to suggest that the marginal Fermi liquid behavior is an evidence of a metallic quantum critical point. 

The three dimensional Hubbard model was studied in the lower doping regime by Schafer et. al \cite{Schafer_etal_2015}. They utilize the dynamical vertex approximation and report findings above the Neel temperature. They concur with our finding that the quasi-particle weight saturates at $N=0.9$. At lower dopings the quasi-particle weight does not saturate but contrasts with our findings in that it does not decrease more rapidly as temperature decreases. 

A follow up point of interest will be to study whether the pseudogap phase, as defined by the partial gap opening or a suppression of the density of state as that in the two dimensional systems, exists for the three dimensional Hubbard model. Thermodynamic quantities could corroborate with the spectral data to support the quantum critical point argument. It has been shown for the two dimensional model that the entropy peaks at the critical doping. \cite{Galankis_etal_2011,Mikelsons_etal_2009} A similar effect should be expected in the three dimensions. As far as we understand, there is no experimental data for the quantum critical point in materials which directly correspond to the Hubbard model. Even for the two dimensional Hubbard model the putative quantum critical point is likely preempted by the pairing instability. However, there is plenty of evidence of a quantum critical point for the Kondo lattice materials \cite{Steppke933,paschen2004hall}. It is a worthwhile direction to study the Kondo lattice model by the dynamical cluster approximation.

\section{Acknowledgement}
We thank Mark Jarrell for his comments and suggestions for this project. This work is funded by the NSF Materials Theory grant DMR1728457. An award of computer time was provided by the INCITE program. This research also used resources of the Oak Ridge Leadership Computing Facility, which is a DOE Office of Science User Facility supported under Contract DE-AC05-00OR22725. Additional support for KMT was provided by the NSFEPSCoR CIMM project under award OIA-154107. 

\bibliography{apssamp}

\end{document}